\documentclass[11pt]{article}
\usepackage{moriond,epsfig}

\def\be{\begin{equation}}
\def\ee{\end{equation}}
\def\bea{\begin{eqnarray}}
\def\eea{\end{eqnarray}}

\begin{document}

\begin{figure}[h]
  \includegraphics[scale=0.32]{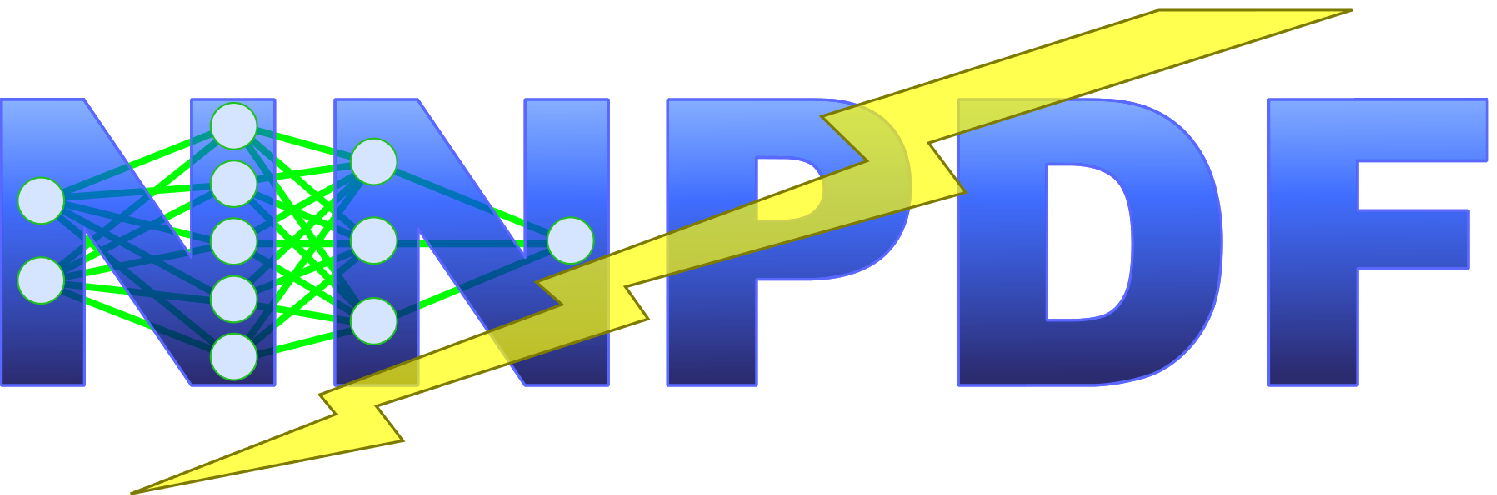}
\end{figure}

\vspace{-2.0cm}

\begin{flushright}
Milano 05/2013
\end{flushright}

\vspace*{4cm}

\title{TOWARDS AN UNBIASED DETERMINATION OF PARTON DISTRIBUTIONS WITH
  QED CORRECTIONS}

\author{Stefano Carrazza}

\address{  
  On behalf of the NNPDF Collaboration\\
  Dipartimento di Fisica, Universit\`a di Milano \& INFN, Sezione di
  Milano\\ Via Celoria 16, I-20133 Milano, Italy
  }

\maketitle

\abstracts{ Electroweak corrections to hadron collider processes
  become relevant at the level of precision reached by present-day LHC
  experiments. We provide a preliminary discussion of the impact of
  electroweak corrections to parton distributions, concentrating on
  electrodynamics corrections to parton evolution equations, and
  showing a preliminary assessment of their impact. Furthermore, we
  determine the parton distribution function of the photon from deep
  inelastic scattering data using the NNPDF methodology.  }

\paragraph{Introduction}

The inclusion of quantum elactrodynamics (QED) corrections to hadron
collider processes, which in turns requires the determination of the
photon parton distribution function (photon PDF) is motivated by the
need for greater precision for LHC measurements, such as the $W$ mass
determination, high mass searches, $WW$
production~\cite{Bozzi:2011ww}. Moreover, a precise determination of
the photon PDF is needed for a reliable computation of several new
physics signals, such as the cross-section for $Z'$ and $W'$
production.
 
Here we will include QED corrections up to leading order (LO) in
$\mathcal{O}(\alpha)$, to next-to-leading order (NLO,
i.e. $\mathcal{O}(\alpha_{s}^{2})$) QCD computations. This choice is
motivated by the similar magnitude of $\alpha_{s}^{2}(M_{Z}^{2})$ and
$\alpha(M_{Z}^{2})$, which suggests that LO QED corrections and NLO
QCD corrections are of a similar size.

Even though our final goal is the computation of LHC processes, we
start by determining the impact of QED corrections and the photon PDF
from deep-inelastic scattering (DIS). In such case, the photon PDF
obtained is determined indirectly by DGLAP evolution, because at LO in
QED there are no photon-induced processes in DIS. Consequently, we
expect the uncertainties of the photon PDF determined from DIS to be
large.  Further information on the photon PDF can be obtained from
collider processes to which the photon contributes at LO, such as
vector boson production. While we will present a first assessment of
the impact of QED corrections on these processes, their inclusion in
the determination of the photon PDF will be presented elsewhere.

\paragraph{PDFs evolution}
\label{sec:evol}

The LO QED evolution equations~\cite{Roth:2004ti,Martin:2004dh} which
are coupled to the standard QCD DGLAP are
\begin{eqnarray}
  Q^{2}\frac{\partial}{\partial Q^{2}}\gamma(x,Q^{2})  =
  \frac{\alpha(Q^{2})}{2\pi}\left[P_{\gamma\gamma}(\xi)\otimes
    e_{\Sigma}^{2}\gamma\left(z,Q^{2}\right)+P_{\gamma
      q}(\xi)\otimes{\sum_{j}}e_{j}^{2}q_{j}\left(z,Q^{2}\right)
  \right],
\end{eqnarray}
\begin{eqnarray}
  Q^{2}\frac{\partial}{\partial Q^{2}}q_{i}(x,Q^{2}) = \frac{\alpha(Q^{2})}{2\pi}\left[P_{q\gamma}(\xi)\otimes e_{i}^{2}\gamma\left(z,Q^{2}\right)+P_{qq}(\xi)\otimes e_{i}^{2}q_{i}\left(z,Q^{2}\right)\right],
\end{eqnarray}
where $\gamma(x,Q^{2})$ and $q_{i}(x,Q^{2})$ are respectively the PDF
of the photon and the $i$-th quark flavor, $P_{ij}(\xi)$ are splitting
functions, $e_{i}$ the quark electric charge, and
$e_{\Sigma}^{2}=N_{c}\sum_{i}e_{i}^{2}$ the sum over all active quark
flavors with $N_{c}=3$. When combining QCD and QED evolution, PDFs
satisfy the momentum sum rule
\begin{eqnarray}
  \int_{0}^{1}dx\,x\left\{
    {\sum_{i}}q_{i}(x,Q^{2})+g(x,Q^{2})+\gamma(x,Q^{2})\right\} =1.
\end{eqnarray}

\begin{figure}
  \begin{centering}
    \includegraphics[scale=0.4]{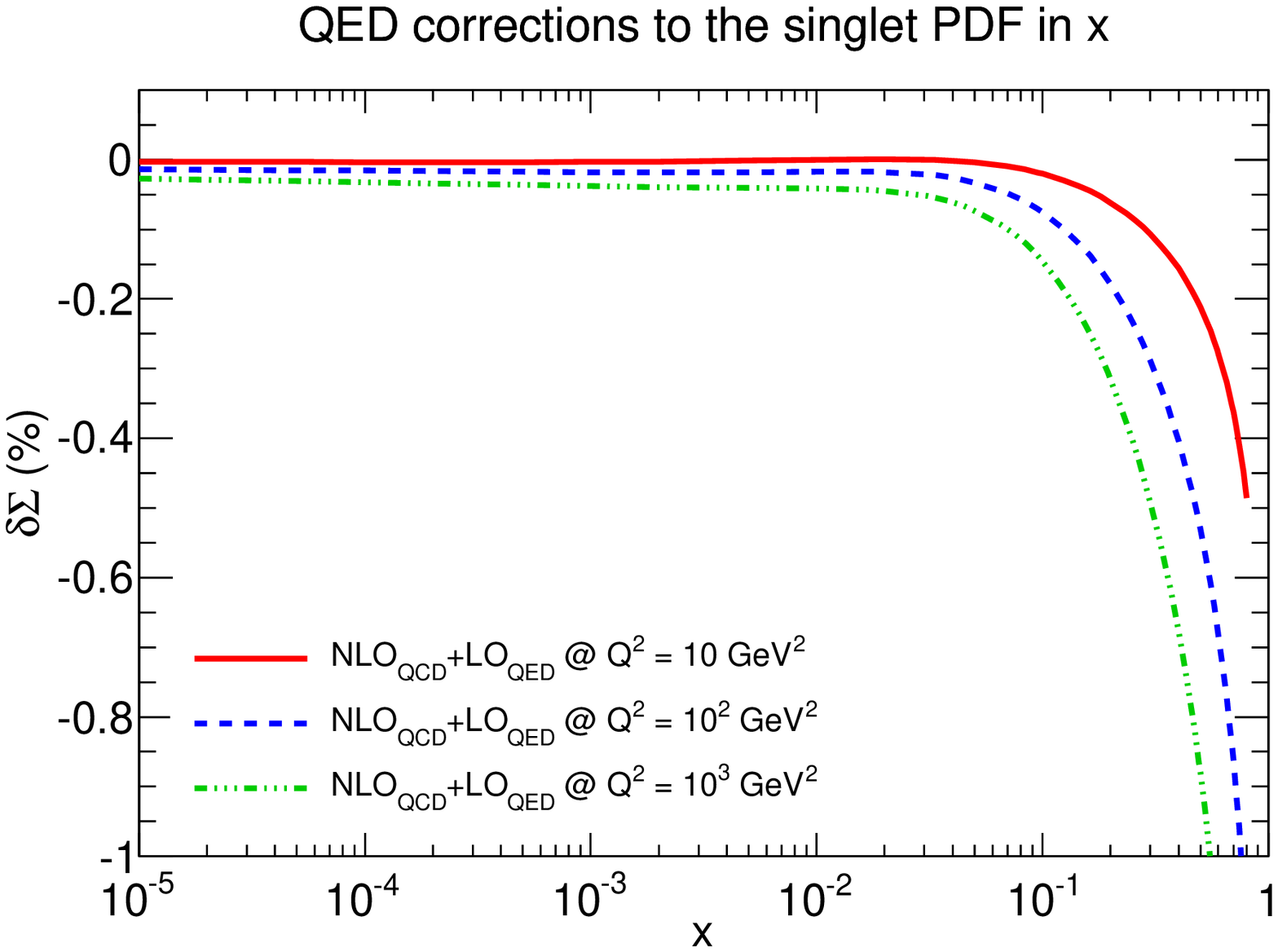}\includegraphics[scale=0.4]{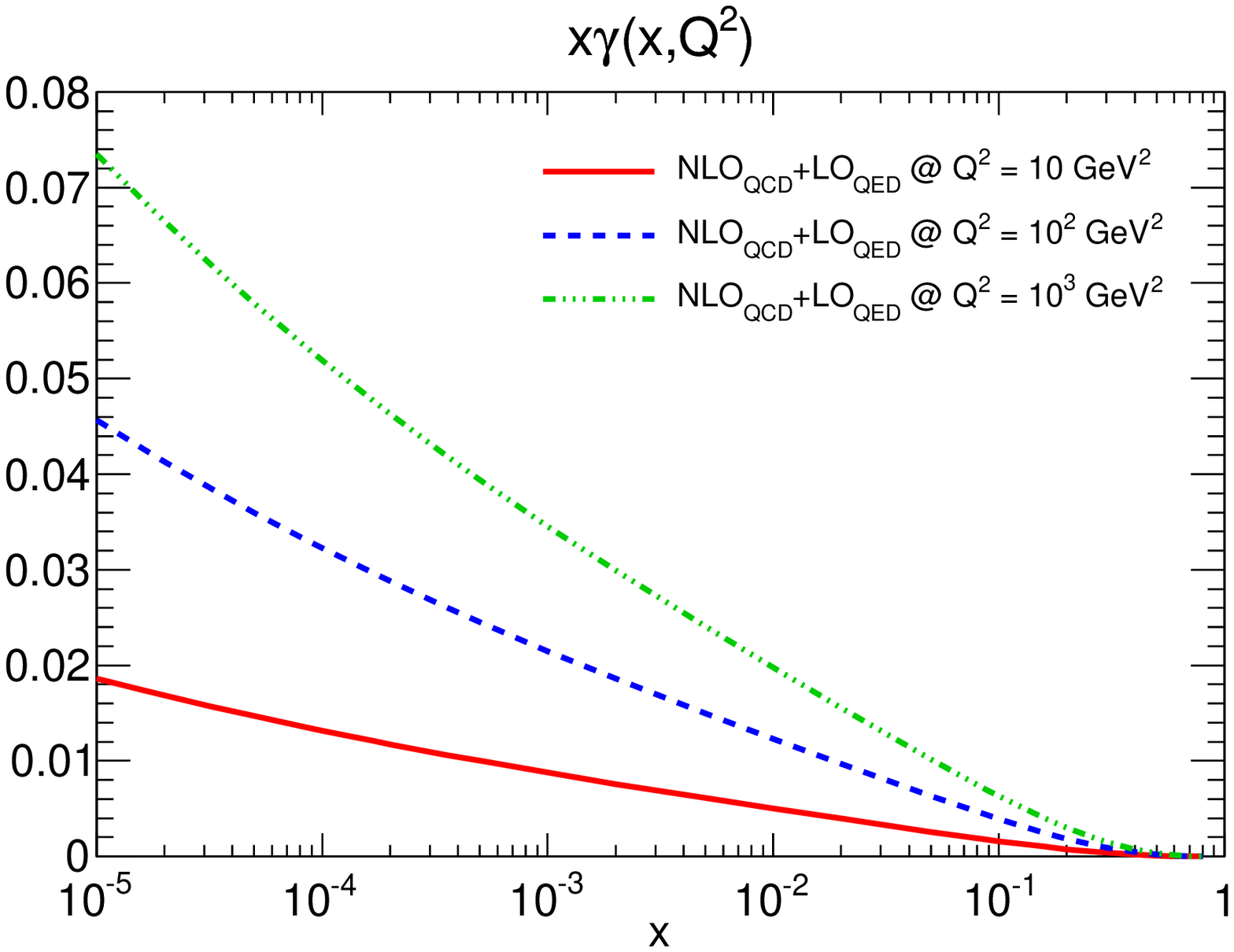}
    \par\end{centering}
  \caption{\label{fig:evolution} Impact of QED
    evolution  on the singlet PDF, $\Sigma =
    \sum_{i}q_{i}$ when
    $\gamma(x,Q^{2}_{0})=0$ (left). The photon PDF generated
    dynamically at higher scales is shown on the right.}
\end{figure}

There are several methods to solve the combined QCD+QED DGLAP
evolution equations, for example by finding a PDF basis which
simultaneously diagonalizes the system as in
Ref.~\cite{Roth:2004ti}. We have developed a combined solution which
optimizes the matching to the NNPDF implementation.

The effect of QED evolution on the singlet PDF is shown in the left
plot of Figure~\ref{fig:evolution}, where the percentage correction to
NLO QCD DGLAP evolution due to the inclusion of QED is shown for
different scales $Q^{2}$, using as input the {\tt Les Houches toy
  PDF}~\cite{Giele:2002hx} and setting $\gamma(x,Q^{2}_{0})=0$ at the
initial scale $Q^{2}_{0}=2$ GeV$^{2}$. We observe differences of the
permille level at large-$x$, though effects are more pronounced at
higher energies: LO QED corrections due to evolution are small,
however the initial photon PDF has been so far assumed to vanish.  The
right plot of Figure~\ref{fig:evolution} shows the photon PDF
generated dynamically at energies above the initial scale $Q^{2} > 2$
GeV$^{2}$.

\paragraph{Photon PDF from DIS data}
\label{sec:disfit}

\begin{figure}
  \begin{centering}
    \includegraphics[scale=0.4]{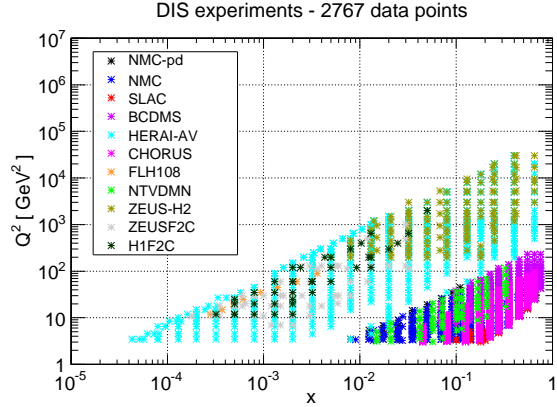}
    \par\end{centering}
  \caption{\label{fig:kinplot} Kinematic coverage of the experimental
    DIS data used in the determination of the photon PDF.}
\end{figure}

With QED corrections to DIS included up to LO through QED evolution,
we can now try to determine the photon PDF from a fit to DIS data.  We
use the DIS dataset included in the NNPDF2.3~\cite{Ball:2012cx}
determination, shown in the $(x,Q^{2})$ plane in
Figure~\ref{fig:kinplot}, which includes 2767 data points.

A consequence of the inclusion of QED corrections  is isospin
symmetry breaking: the neutron PDFs cannot no longer be obtained from the proton
PDFs  using isospin, i.e. $u^{n} \neq d^{p}$, $d^{n} \neq
u^{p}$. However, we assume
that isospin holds at the starting scale $Q^{2}_{0}$. It is then broken
dynamically by  DGLAP evolution.

In the NNPDF framework~\cite{Ball:2012cx,Ball:2011uy}, we generate a
set of Monte Carlo replicas of the data, then from each replica we
extract a PDF set.  The minimization is performed by a genetic
algorithm and the best fit is determined by cross-validation.  One of
the most important advantages of this methodology is that it minimizes
the bias related to the choice of functional form of PDFs.  We
parametrize the photon PDF, like all other PDFs, with a feed-forward
neural network with 2-5-3-1 architecture, which corresponds to a total
of 37 parameters to be determined during the minimization
procedure. Positivity of the photon PDF at LO is imposed by squaring
the output of the neural network. The rest of the procedure is the
standard NNPDF one.

\begin{figure}
  \begin{centering}
    \includegraphics[scale=0.4]{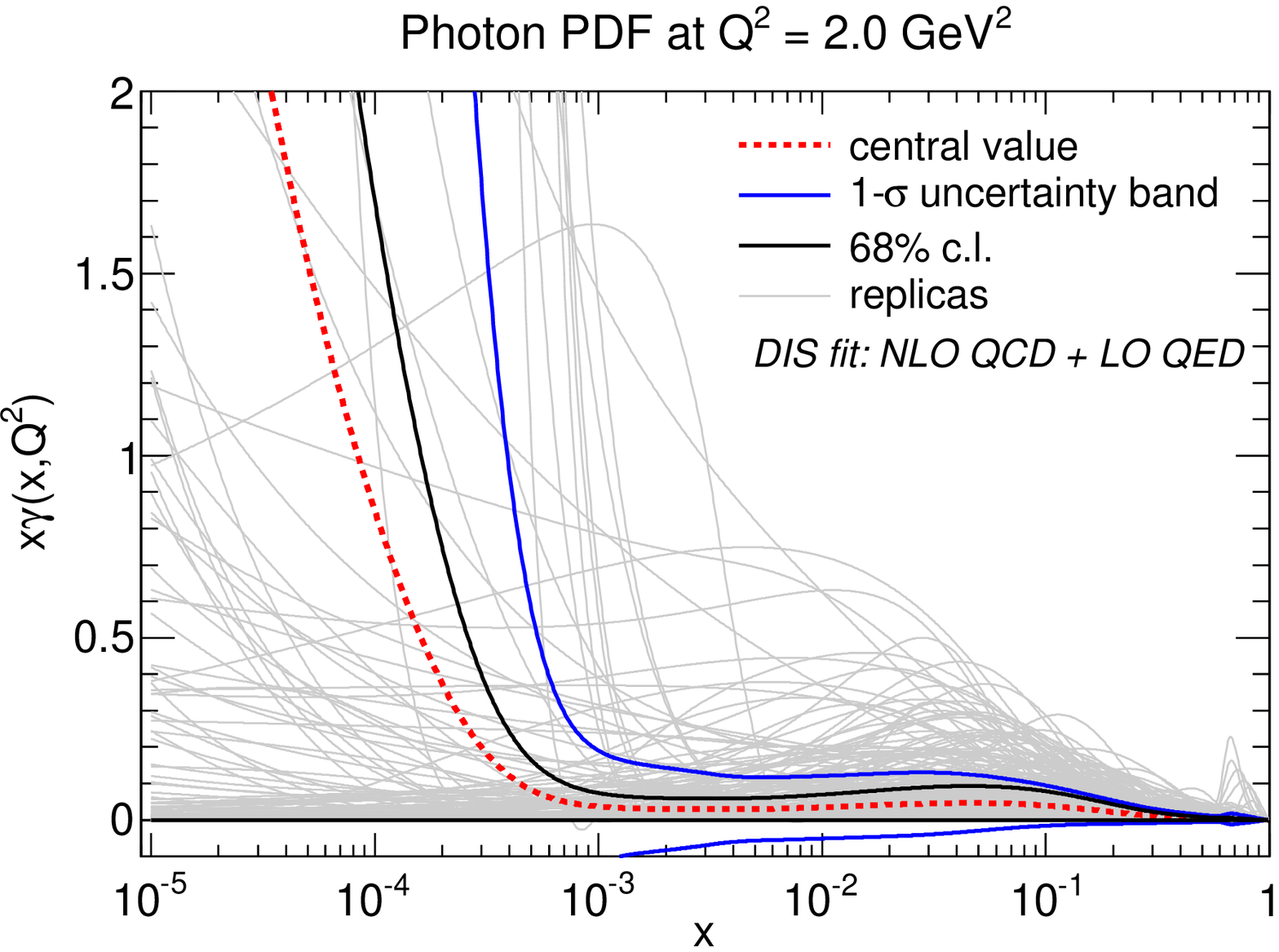}\includegraphics[scale=0.4]{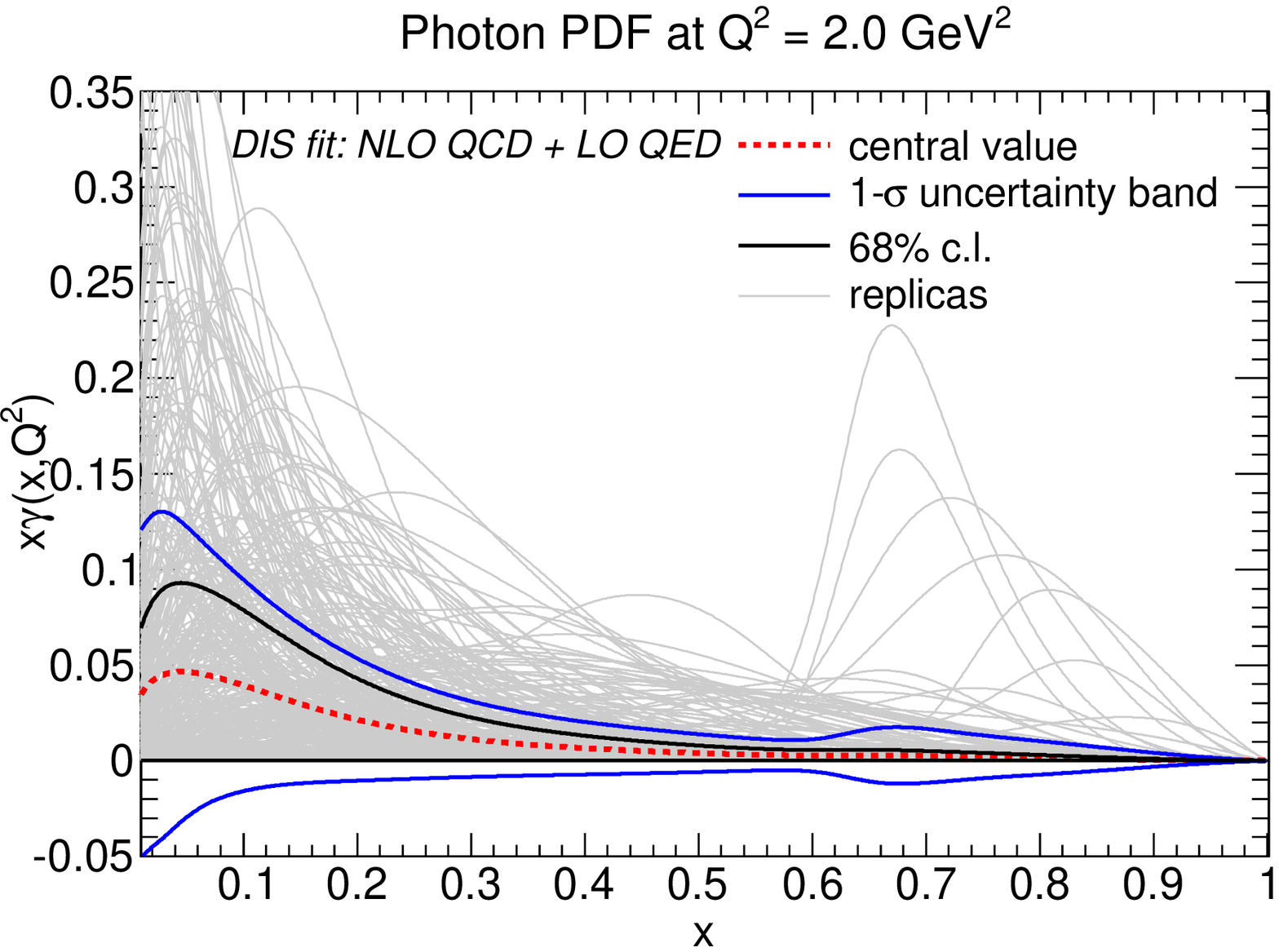}
    \par\end{centering}
  \caption{\label{fig:photon}The photon PDF extracted from DIS data,
    NLO QCD + LO QED, at $Q^{2}=2$ GeV$^{2}$. The Monte Carlo PDF set
    includes
500 replicas.}
\end{figure}

The photon PDF from DIS is showed in Figure~\ref{fig:photon}, in
logarithmic (left) and linear (right) scales. The plots show the
central value, the 1-$\sigma$ and the 68\% c.l. uncertainties bands,
defined around the mean value, and the 500 PDF replicas. The photon
from DIS data is compatible with zero with large uncertainties. The
photon PDF is less uncertain at central and large-$x$ than at
small-$x$, due to the lack of data points in this region (see
Figure~\ref{fig:kinplot}). The overall fit quality, as measured by the
total $\chi^{2}$ per data point, is $\chi^{2}=1.10$.

\paragraph{The impact of the photon PDF from DIS}

In order to understand the impact and the quality of the photon PDF
just presented, we have computed $Z\rightarrow\mu^{+}\mu^{-}$
production in proton-proton collision at $\sqrt{s} = 14$ TeV with
$|\eta^{l}| \leq 2.5$ and $p^{l}_{T} \geq 20$ GeV using {\tt
  HORACE}~\cite{CarloniCalame:2007cd}, which is a Monte Carlo event
generator for Drell-Yan processes including the exact 1-loop
electroweak radiative corrections $\mathcal{O}(\alpha)$. {\tt HORACE}
also provides the possibility to complement the $\mathcal{O}(\alpha)$
with photon initiated (photon-induced) processes at Born and NLO
levels. The Born photon-induced contribution must be included at the
order at which we are working. We have also included the NLO
corrections, which however have a very small effect.

\begin{figure}
  \begin{centering}
    \includegraphics[scale=0.4]{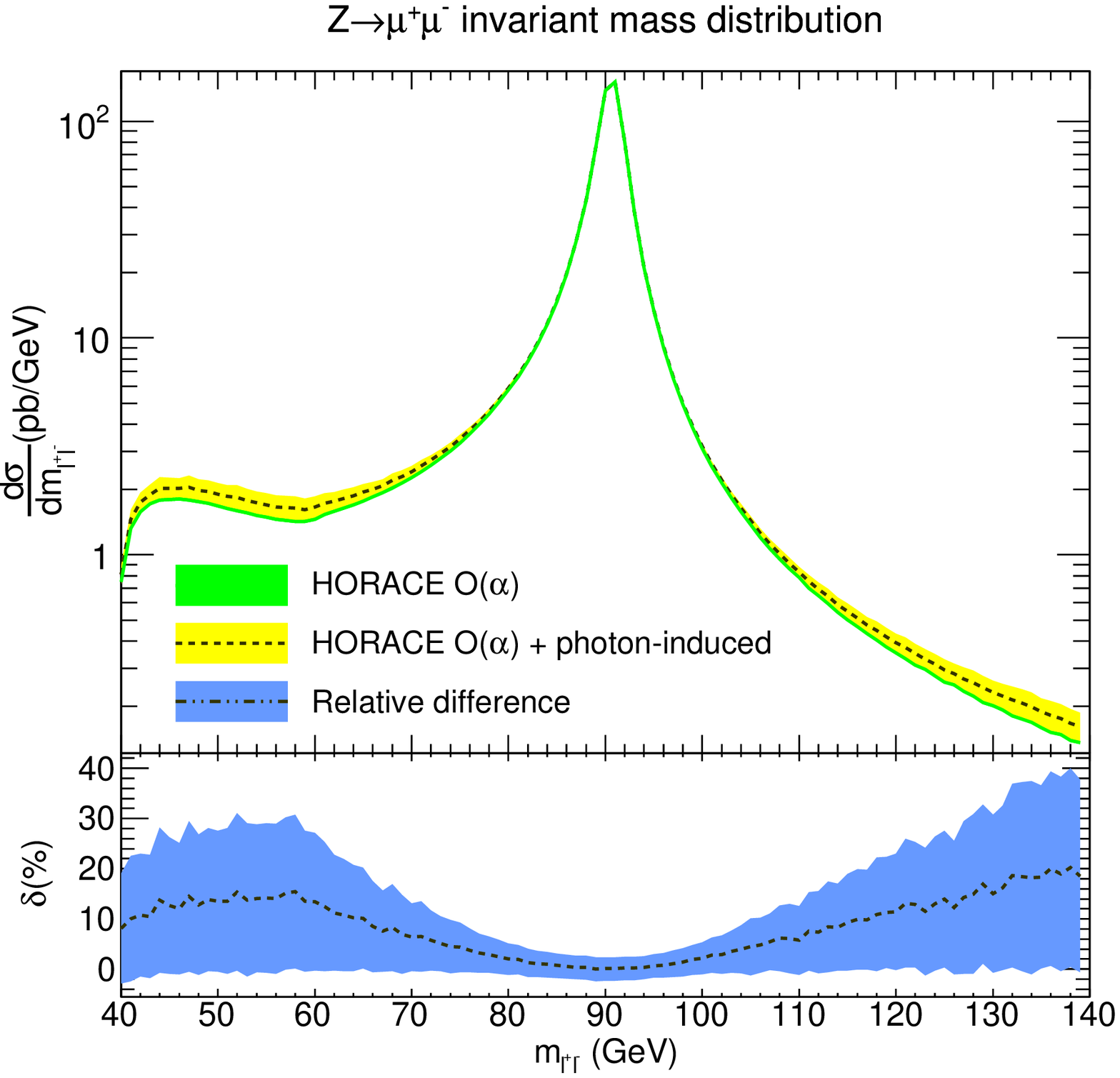}\includegraphics[scale=0.4]{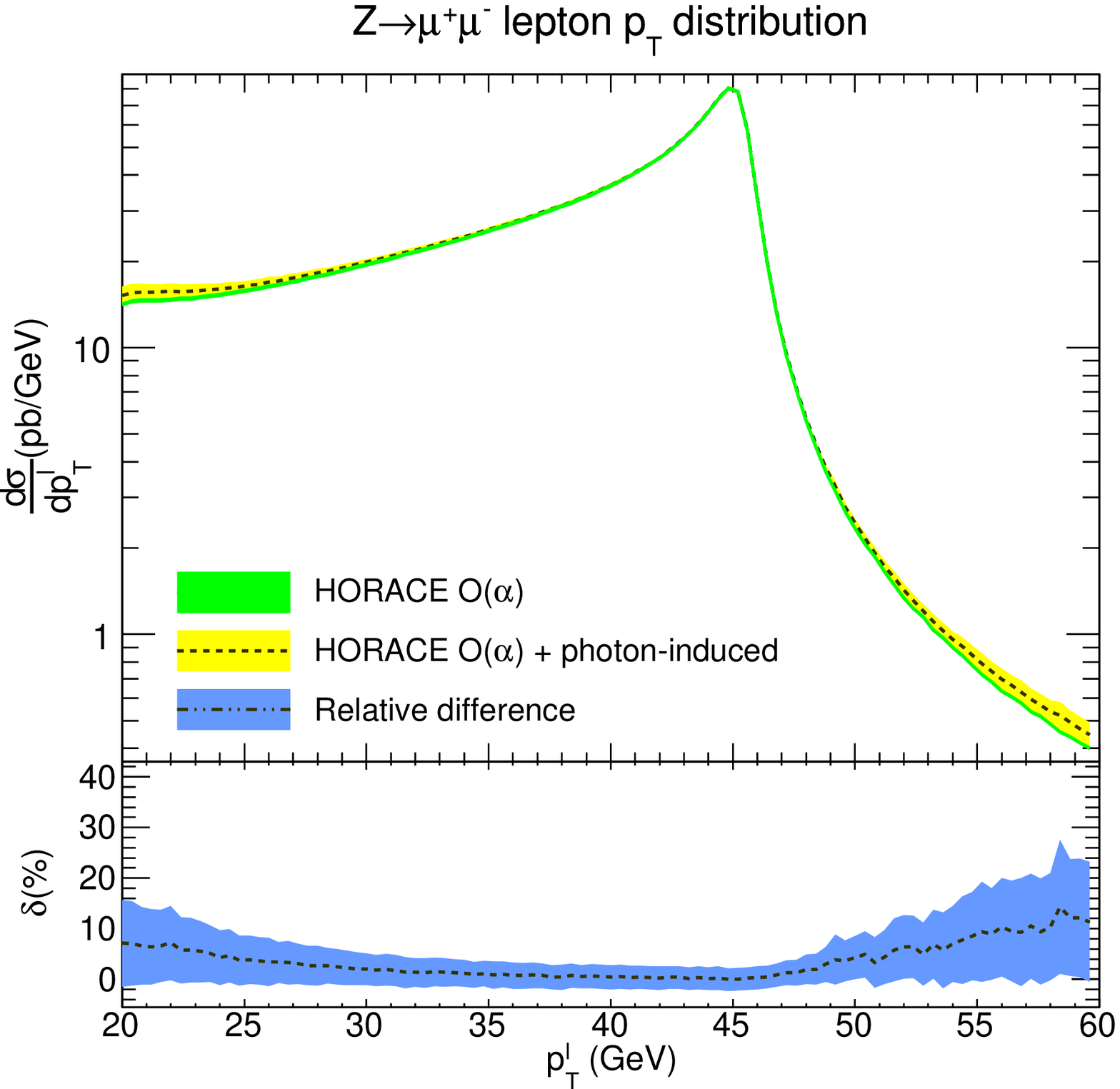}
    \par\end{centering}
  \caption{\label{fig:horace} $Z$ invariant mass and lepton
    $p_{T}^{l}$ distribution computed using {\tt HORACE} at
    $\sqrt{s}=14$ TeV, with $|\eta^{l}| \leq 2.5$ and $p^{l}_{T} \geq
    20$ GeV. Results to $\mathcal{O}(\alpha)$ with and without
    photon-induced contributions are compared.}
\end{figure}

Figure~\ref{fig:horace} shows the $Z$ invariant mass and the lepton
$p_{T}^{l}$ distributions, for 100 replicas and using the 68\%
c.l. uncertainty band. There is a moderate relative difference between
$\mathcal{O}(\alpha)$ and $\mathcal{O}(\alpha)$ + photon-induced
processes due to the photon PDF, in the region of the $m_{ll}$ and
$p_{T}^{l}$ peaks, which rapidly increases when going away from the
respective peaks. The increase of the central value is expected, in
fact, for example at the Born level we are adding the $\gamma\gamma
\rightarrow \mu^{+}\mu^{-}$ processes to the usual $q\bar{q}
\rightarrow \mu^{+}\mu^{-}$, however also the uncertainties grow when
going far from the peak region. These large uncertainties reflect the
uncertainty in the photon PDF, which is insufficiently constrained by
DIS data.

\paragraph{Outlook}

Our results show that the photon PDF is only poorly constrained by DIS
data, especially at small-$x$. Corrections to $Z$ production due to
photon-induced processes in the presence of such a large unconstrained
PDF off the peak become rapidly larger than the data allow. This
suggests that these and similar data should be used to determine the
photon PDF itself. This determination will be presented elsewhere.

\end{document}